# An IDS scheme against Black hole Attack to Secure AOMDV Routing in MANET

[1]Sonal Shrivastava, [2]Chetan Agrawal & [3]Anurag Jain

[1]M-Tech (CSE), [2]Asst Prof (CSE) & [3]HOD (CSE)
Radharaman Institute of Technology & Science Bhopal, India
[1]sonal.27486@gmail.com,[2]chetan.agrawal12@gmail.com
[3]anurag.akjain@gmail.com

## *Abstract*

*In Mobile Ad hoc Network (MANET) all the nodes are freely moves in the absence of without ant centralized coordination system. Due to that the attackers or malicious nodes are easily affected that kind of network and responsible for the routing misbehavior. The routing is network is mandatory to deliver data in between source and destination. In this research we work on security field in MANET and proposed a novel security scheme against routing misbehavior through Black hole attack. The Ad hoc On demand Multipath Routing (AOMDV) protocol is consider for routing and also to improves the routing quality as compare to single path routing protocol. The attacker is affected all the possible paths that is selected by sender for sending data in network. The malicious nodes are forward optimistic reply at the time of routing by that their identification is also a complex procedure. The proposed Intrusion Detection System (IDS) scheme is identified the attacker information through hop count mechanism. The routing information of actual data is reached to which intermediate node and the next hop information is exist at that node is confirm by IDS scheme. The black hole attacker node Identification (ID) is forward in network by that in future attacker is not participating in routing procedure. The proposed security scheme detects and provides the deterrence against routing misbehavior through malicious attack. Here we compare the routing performance of AOMDV, Attack and IDS scheme. The performance of normal multipath routing and proposed IDS scheme is almost equal. The attacker has degrades the whole routing performance but observed that in presence of attacker, routing misbehavior is completely block by the proposed IDS scheme and recovers 95 % of data as compare to normal routing.*

# I. Introduction

A mobile ad hoc network (MANET) is a network consisting wireless mobile nodes that conversation with each other without centralized control or established infrastructure [1]. These nodes which are inside each and every one radio range can convey precisely, while distance nodes count on their neighboring nodes to forward packets. In MANETS every node can be a host or router. Mobility, an advantage of wireless communication, allows a freedom of moving around although being linked to a network environment. Ad-hoc networks are so adaptable that nodes can join and move a network easily as compare to wired network [2]. Such networks can be used in the battlefield application, in disaster management and in remote areas where establishment and management of fixed network is not possible. These can also be used in the areas where the establishment of fixed infrastructure is very difficult. MANETs can also be used to deploy and coordinate the drones in the battle field.

But this flexibility of mobile nodes proved in an ever-changing topology that forms it very tough in establishing secure ad-hoc routing protocols. The radio channel employ for ad hoc networks is propagate in nature and is common by all the nodes in the network. Data transmitted by a node is collect by all the nodes within its direct transmission range. So attackers can easily affect the data being transmitted in the network. If the ad hoc network lacks some form of network level or link layer security, a MANET routing protocol [3] will be more vulnerable to many forms of malicious attacks. It can be simple attack like snooping network traffic, transmissions replay, manipulation of the packet headers, and redirecting the routing messages, within an Ad hoc network without any appropriate security provisions. In Black hole attack, malicious nodes get a chance to attack during route discovery process. A black hole means that one malicious node apply the routing protocol to affirm itself of having shortest path to the destination node, and drops routing packets and does not send packets to its adjacent node [4]. A single Black hole node can easily attack on mobile Ad hoc networks [5]. There is various detection schemes for detecting single black hole, but failed when cooperative black hole attack occurs. Cooperative black hole attack means malicious nodes act in a group. In this attack, one malicious node receives data and forwards it to other malicious node instead of forwarding it towards destination. To provide complete security in this paper we proposed a secure IDS scheme for AOMDV [6] routing protocol of MANET. We required secure routing procedure protocol for appends the possibilities of actual data delivery in network as compare to attacker infected data. Since nodes involved in the routing cannot by themselves ensure the secure and uninterrupted delivery of transmitted data.

The absence of centralized authority is the major reason of attack in MANET. The malicious nodes are easily affected the routing performance to be a part of it. It implies malicious node are already present in network and always perform the same function as normal nodes, hence the identification of these nodes are not easy.

The routing misbehavior of attacker or malicious nodes drop the data routed through that nodes or range on these nodes. Misbehaving nodes can interrupt the route discovery by acting the destination, by replying with stagnant or demoralized routing information, or by propagating progressed control traffic. With the substantial thing in secure routing in the mobile

Ad-hoc network we propose a new algorithm for developing network interpretation from attack like throughput, Routing load and delay [8].

Security in Mobile Ad-Hoc Network (MANET) is the greatest essential firm for the basic affinity of network. Accessibility of network services, esoteric and integrity of the data can be accomplished by ensuring that security issues have been met. MANET frequently suffer from security attacks due to its features like unclosed medium, changeful its topology dynamically, absence of central monitoring and management, cooperative algorithms and no clear defense mechanism. These aspects have developed the battle field position for the MANET against the security threats.

The remaining of this paper is organized as follows. In Section II we show the typed of attack classification. The Section III is presents the Previous work done in this field. Section IV is described the proposed IDS security scheme in depth. The Section V presents the description of simulation environment. The section VI is presents the simulation results using ns2 are presented and In Section V, we conclude our work with future extension.

## II. Types of attacks in MANET

The attacks in MANETS are classified into two major categories [7, 9], namely

**Passive Attacks -** Passive attacks are those, established by the adversaries merely to intrude the data exchanged in the network. These attackers in any mode don't divert the action of the network. Such attacks recognition get very rigorous since network itself does not impacted and *they* can depreciate by using persuasive encryption techniques.

**Active Attacks -** The active attack attempts to change or demolish the data that is being exchanged, by that distressing the standard operation of the network. In MANET malicious and unwanted nodes to interrupted the natural operating in the network.

Some of the common attacks [9, 10] in MANET are as follows:-

1) **Black Hole Attack -** The aim of this attack is to enhance the congestion in network. In this attack the adversary node does not transmit any packets transmitted to it, rather drops them all. Due to this attack the packets sends by the nodes do not come their proposed destination and the congestion in the network expand due to retransmissions.

2) **Wormhole Attack -** In wormhole attack, two concurring nodes are mentioned and one node tunnels the packet to some other node in the consistent network over a huge speed private wired link or wireless link these packets are then dislike from that location into the network. This tunnel between two selfish nodes is known as wormhole.

3) **Denial of Service (DoS) Attack and Flooding -** The purpose of this attack is to exhaust the gentle working of the network. This attack is executed by constantly forwarding packets into the network inducing the proposed node in the network to work them and hold them employed resulting in the colliding of that node. By executing this attack, the attacker holds the targeted node active in developing its induced packets and discarding the legitimized RREQs to be dropped. This attack can get the network substructure to break-down.

4) **Sybil Attack -** In the Sybil attack [11], an attacker act to have multiple identities. A malicious node can play as if it were a enormous number of nodes either by acting other nodes or justly by claiming wrong identities. Sybil attacks are divided into three divisions like direct or indirect communication, fabricated or stolen identity, and simultaneity. In the direct communication, Sybil nodes convey instantly with legitimize nodes, whereas in the indirect convey messages sent to Sybil nodes are routed through malicious nodes. An attacker can assemble a new identity or it can simply rob it after destructing or temporarily disenabling the impersonated node. All Sybil identities can act together in the network or they may be cycled through.

**5) Byzantine attack -** In a Byzantine attack [12] compromised intermediate node perform by itself, or a set of compromised intermediate nodes execute in complicity and perform attacks such as generating routing loops, transmitting packets through non-optimum route, or selectively dropping packets, which outcomes in annoyance or degradation of the routing services.

**6) Rushing attack -** In rushing attack [7, 13] an attacker occupy the RREQ packet from source node disperse over the packet promptly to all the other nodes in the network, earlier they get the same packet from the source. Formerly the primary RREQ packet comes to the nodes, they consider it is a replicate and repudiate it since they preliminarily have the packet from adversary.

## III. Work Accomplished in this field

The Secured routing schemes for MANETs have received increasing attention in the recent years, with main focus on data forwarding**.**

In this paper [14], an approach have been proposed to diminish the black hole attack using AOMDV (Ad hoc on Demand Multipath Distance Vector) routing protocol. Some recoveries have been made in AOMDV protocol. The proposed approach based upon AOMDV (Ad hoc on demand Multipath Distance Vector) route discovery and creates a new logic for avoiding black hole attack using Legitimacy or trust factor of a node. This approach detects black hole node and discovers node disjoint multipath, which reduces the overhead of a specific node. These meliorate produce the protocol vigorous against black hole attack along multipath route discovery process.

In this paper [15] they proposed SAODV protocol which is the extension of AODV. This protocol is secure and efficient MANET routing protocol which aims to address the security is not strong of the AODV protocol and is able to perform of enduring the black hole attack.

This paper [16] proposed a secured message security scheme for MANETs (our so-called T-AOMDV) that uses a trust-based multipath AOMDV routing combined with a soft-encryption methodology to securely transfer messages. More precisely, our approach consists of three steps: (1) Message encryption: where at the source node, the message is segmented into three parts and these parts are encrypted using one another using some XOR operations, (2) Message routing: where the message parts are routed separately through different trust based multiple paths using a novel node disjoint AOMDV protocol, and (3) Message decryption: where the destination node decrypts the message parts to recover the original message.

In this paper [17], they proposed a method uses Intrusion Detection using Anomaly Detection (IDAD) to defend against black hole attacks established by both single and multiple black hole nodes. It proved the specific result increases network performance by reducing formation of control (routing) packets including effectively defend black hole attacks opposed to mobile ad-hoc networks.

In this paper [18], they proposed a method uses promiscuous mode to find malicious node and transmit the data of malicious node to every some other nodes in the network. The efficiency of suggested mechanism as throughput of the network does not decay in existence of the black holes.

In this paper [19], they proposed two possible solutions to study black hole attack. The first solution is to study several route to the destination. The second is to apply the packet sequence number contained in any packet header. In study to AODV routing scheme, the second solution is superior and of the route to the destination rely upon on the pause time at a lowest cost of the delay in the networks.

In this paper [20], they have proposed a solution the requesting node wait and check the replies from all neighboring node to find a safe route. It is provide better performance than the conventional AODV in the existence of Black holes with smallest additional delay and overhead.

In this paper [21], they apply a reactive routing protocol called as Ad hoc On-demand Distance Vector (AODV) routing for examine of the outcome of the black hole attack when the destination sequence number is altered via simulation. Then, they determine characteristic in order to define the normal state from the character of black hole attack. They proposed training scheme for huge accuracy detection by modifying the training data in every given time intervals and adaptively specifying the normal state according to the changing network environment.

## IV. Problem Statement

In MANET environment the problem is, nodes are supposed to collaborate among each other dynamically to give routing service and transmit packets. This need represent a security challenge when malicious nodes are exhibit in the network. Certainly, the presence of such nodes may not simply interrupt the normal network operations, but cause serious message security concerns. The security concern is necessary in absence of centralized administration because of no system in network is to monitor the routing information and malicious activities. The routing misbehavior is degrades the network performance by dropping the data packets or capturing the data packets in network.

1. The attacker replies false route information.
2. The all data is dropped by attacker to forward through malicious nodes.
3. The sender has continuously tried to send data after failure number of iterations.
4. The routing packets are more deliver to send little amount of data.

The proposed scheme is improves the routing misbehavior from malicious nodes

## V. Proposed IDS Scheme to secure MANET

Route discovery is a susceptibility of on demand Multipath Ad-hoc routing protocols, especially (Ad hoc On Demand Multipath Distance Vector) AOMDV, which an adversary can get to act a black hole attack on mobile ad-hoc networks. A malicious node in the network evolving an RREQ message respond to source nodes by forwarding a fake RREP message that includes valuable parameters to be preferred for packet delivery to destination nodes. After reassuring (by sending a fake RREP to affirm it has a route to a destination node) to source node that it will transmit data, a malicious node begins to drop all the network traffic it.

Secure connection has been established between source nodes to destination nodes. The Proposed IDS algorithm finds out the multiple routes from source to destination using AOMDV routing algorithm. After finding multiple routes, all the routes are classified based on the conviction factor of existence of attacker in route. Then it will prefer the best route which is having no existence of malicious attack or black-hole attack. In this method the data is secured in presence of IDS algorithm. Then it routes the data through best single route of multiple established path in MANET.

In this research we consider three modules of routing:

**a) Normal AOMDV routing module:** To evaluated the performance of normal Multipath routing protocol without presence of attacker.

**b) The second in presence of black hole attack: -** The multipath is reliable for communication because the attacker blocks the established route then it delivers data through next possible routes but multiple attackers are blocks all the possible paths.

**c) The third is proposed IDS module:** This module is proposed to provide security in presence of black hole attack. The attackers are absolutely performing no routing misbehavior and provide reliable routing. The whole procedure of IDS algorithm is mentioned in next point.

**Proposed Algorithm to Identify and Prevent from Attack:**

Number of nodes = 50

Routing Protocol = AOMDV

Type of attacker = Black hole as a Malicious attacker

Security Provider = IDS (Intrusion Detection System)

**Step1:** Sender has sending the request to all intermediate nodes between sources to destination.

**Step2:** Add the next hop in routing table if we have to destination route, otherwise rebroadcast the request and maintaining the hop count information.

**Step 3:** If destination is found then select the route of minimum hop count and deliver data through that minimum hop count path h.

1. Multiple paths are selected on the basis of hop counts $h_{1,} h_{2,} h_{3 \ldots\ldots}$ $h_n$, n=1,23…

2. $\sum$ Hn = ($h_{1,} h_{2,} h_{3 \ldots\ldots}$ $h_n$) up to destination is Minimum then select for data sending and next route of hop count $h_{1,} h_{2,} h_{3} \ldots\ldots h_n \geq$ Min is select for multiple path.

**Step4:** We compare the AOMDV routing table through IDS system to next hop routing table, if table is matched it means no attack in the network and route is true, and then forward all data packet.

**Step5:** If next node is false, and the next hop information is not matched (M means data entry)

If next hop $h1_{,} h_{2,} h_{3 \ldots}$ $h_{n-1} \neq$ M.

It means no previous data deliver through that hop, insert the table new entry which have shortest path to destination.

**Step6:** If next hop is true, sending data through that hop is false then send the data packet for checking the reliability through proposed IDS security scheme.

**Step7:** IDS (Intrusion Detection system) verify if routing table information is not matched related to actual hop count means some misbehavior activity occurs in the network through malicious nodes.

**Step8:** Applied prevention scheme is and block that hop and change the path, forward data packet. Also forward the nodes ID (identification) in network by that the attacker neither is nor select in routing procedure.

**Step 9** If the attacker is be present in selecting path for data delivery then avoids that path and preferred another suitable path from multiple paths established by AOMDV.

Attacker exists on Hop count $h_{1,} h_{2,} h_{3} \ldots\ldots h_n$ = Min then,

Select route of Hop count $h_{1,} h_{2,} h_{3} \ldots\ldots h_n \geq$ Min

**Step10:** If routing is matched then forward data packet until send all data packet reach to destination.

**Step 11:** Exit

**Flow Chart of Proposed IDS Security Scheme against Malicious Attack**

The flow of proposed security scheme is represents the steps to identified the attacker and obstruct the activities of attacker by that secure communication in possible.

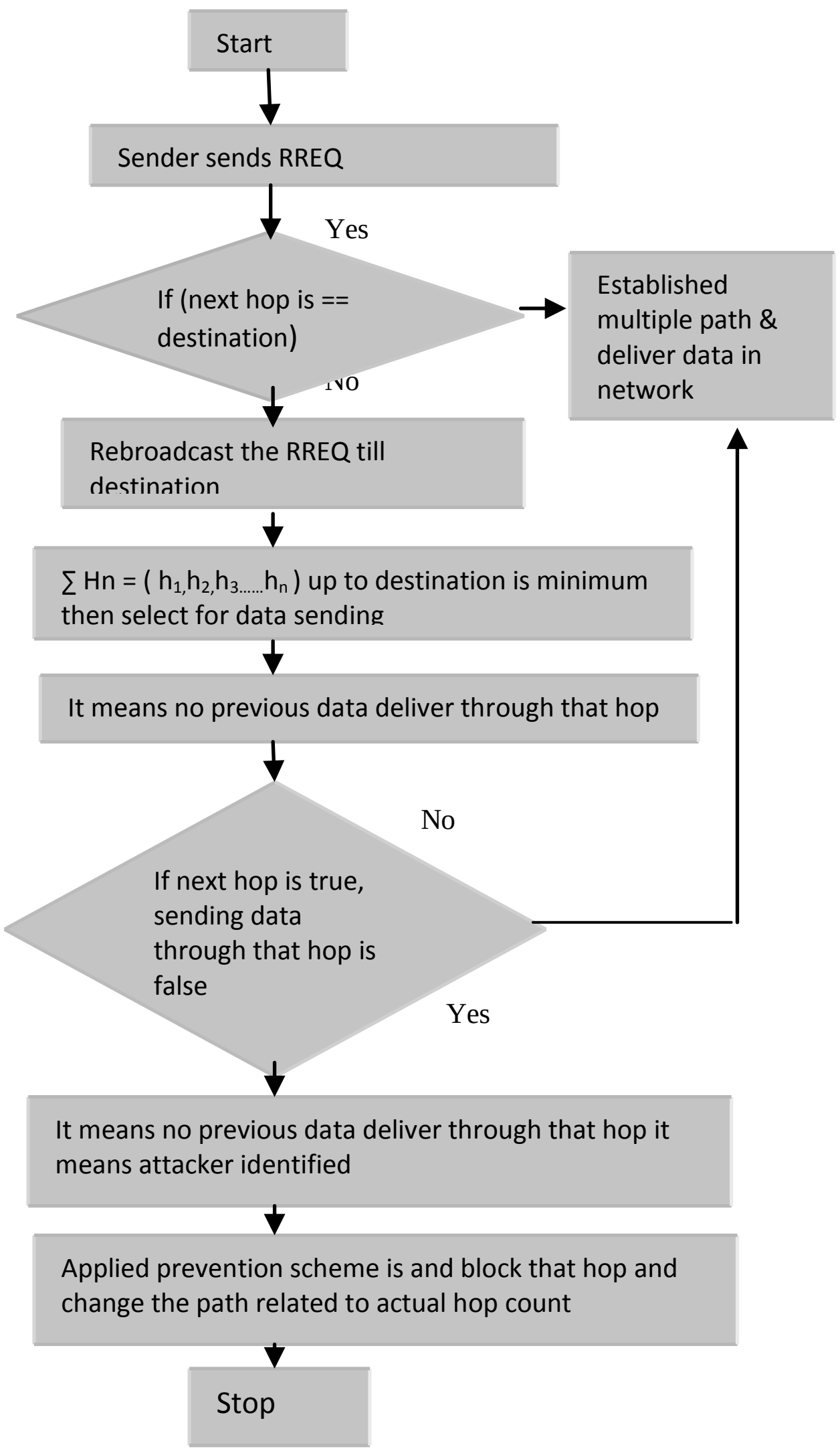

**Fig.1. shows that proposed flow chart of algorithm**

**Hop count** –The hop count represents the total number of devices of data packet passes between sources to destination. The more hops data must traverse to reach their destination, the greater the transmission delay incurred.

**Routing Table -**A routing table contents the data essential to transmit a packet along the finest route toward its destination. Each packet includes data about its source and destination. When a packet is collected, a network tool observes the packet and equals it to the routing table entering giving the best equivalent for its destination. The tables then supply the tool with instructions for forwarding the packet to the next hop on its path across the network.

**IDS** –Intrusion Detection system (IDS) is the process of detecting an adversary and preventing its subsequent action. It is anomaly activities will monitor network traffic and compare normal activities

## V. Simulation Description & Parameters

The simulation of all three modules i.e. normal AOMDV routing, Attack in AOMDV and IDS scheme against Malicious attack in AOMDV done in Network Simulator (NS-2) version 2.3. 1

A network was creating for the simulation aim and then observed for a number of parameters. The TCL (Tool Command Language) of modules simulated for 20, 40, 60, 80 and 100 nodes. Simulation time is taken 100 sec. Each node moves randomly and has a transmission range of 250m. The minimum speed for the simulations is 3 m/s while the maximum speed is 30 m/s. Each mobile node in the MANET is allotted primary location within the simulation dimensions of 800×800 meters and joins the network at an arbitrary time. The packets are created using FTP and CBR with rate of 3 packets per seconds. Nodes are generally allocated when initiated, and the original location for the node is defined in a movement scenario file generated for the simulation using a factor inside ns-2. The propagation model is used two ray ground and the MAC layer technology of 802.11 is considered for wireless communication. The number of attacker nodes is created 4 and against them IDS nodes are plot 2 in network.

### Performance Metrics

The following performance metrics are used for comparing the performance of three modules:

**i. The packet delivery ratio -** The ratio of the data delivered to the destination to the data sent out by the source. The PDF shows how successful a protocol performs delivering packets from origin to destination.

**ii. The average end-to end delay -** This is the average delay between the sending of packets by the source and its receipt by the receiver. This includes all possible delays reasons during data gaining, route discovery, processing at intermediate nodes, retransmission delays, and propagation time. It is measured in milliseconds.

**iii. The normalized routing overhead -** The number of routing packets transmitted per data packet delivered at the destination. The routing overhead minimum is shows better performance.

**iv. Throughput** - Throughput is the average rate of successful message delivery over a communication channel. A high throughput network is desirable.

## V. EVALUATED RESULT

The results evaluated on the basis of considered simulation parameters are mentioned in this section.

**a). Data Drop Analysis in case of AOMDV, Attack and IDS -** The packet drop due attack at the time of routing is the routing misbehavior in network. The attacker is affected the performance of routing by dropping the data packets in network. In this graph the packet drop percentage is identify in case of routing misbehavior of attacker. The performance is illustrated in nodes density of 20, 40, 60, 80 and 100. The drop percentage is identified from the trace file and in trace file only attacker nodes are drop the packets. The drop percentage is about 19% at the end of simulation. The drop through attacker is not identified in presence of proposed IDS scheme. The security scheme obstruct the attacker activities and provide attacker free network. That’s why the drop through in case of IDS scheme in not identified in IDS module simulation.

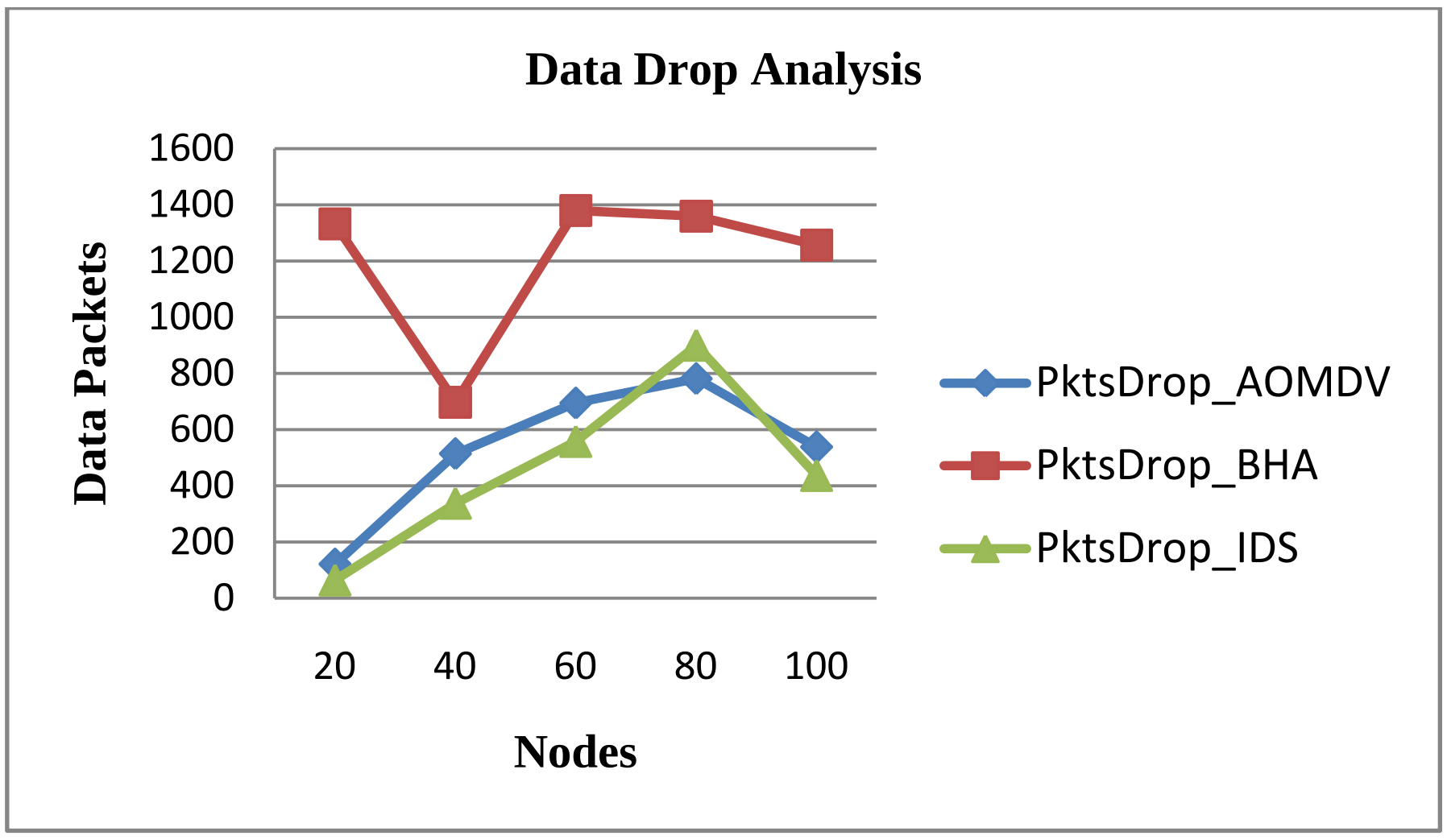

**Fig. 1 Data Drop Analysis**

## b). Packet Delivery Ratio Analysis in case of AOMDV, Attack and IDS

The packets successful transmission is improves the performance of network besides that the packet dropping is degrades the performance of network. The routing misbehavior through black hole attack is degrades the percentage of data receiving in node densities of 20, 40, 60, 80 and 100. The attacker is consuming whole data packets that are not forwarded to destination after positive route reply. The percentage of packets successful receiving in case of normal AOMDV, Attack and IDS scheme is illustrated in this graph. The attacker performance is about 2 % up to simulation time of 50 seconds. The attacker has drop the most of the data packets by that the routing performance of AOMDV routing is degrades. The proposed IDS security scheme is improve the PDR performance is presence of attacker. The PDR in case of IDS scheme is about 95 % and it is almost equal to normal AOMDV routing performance.

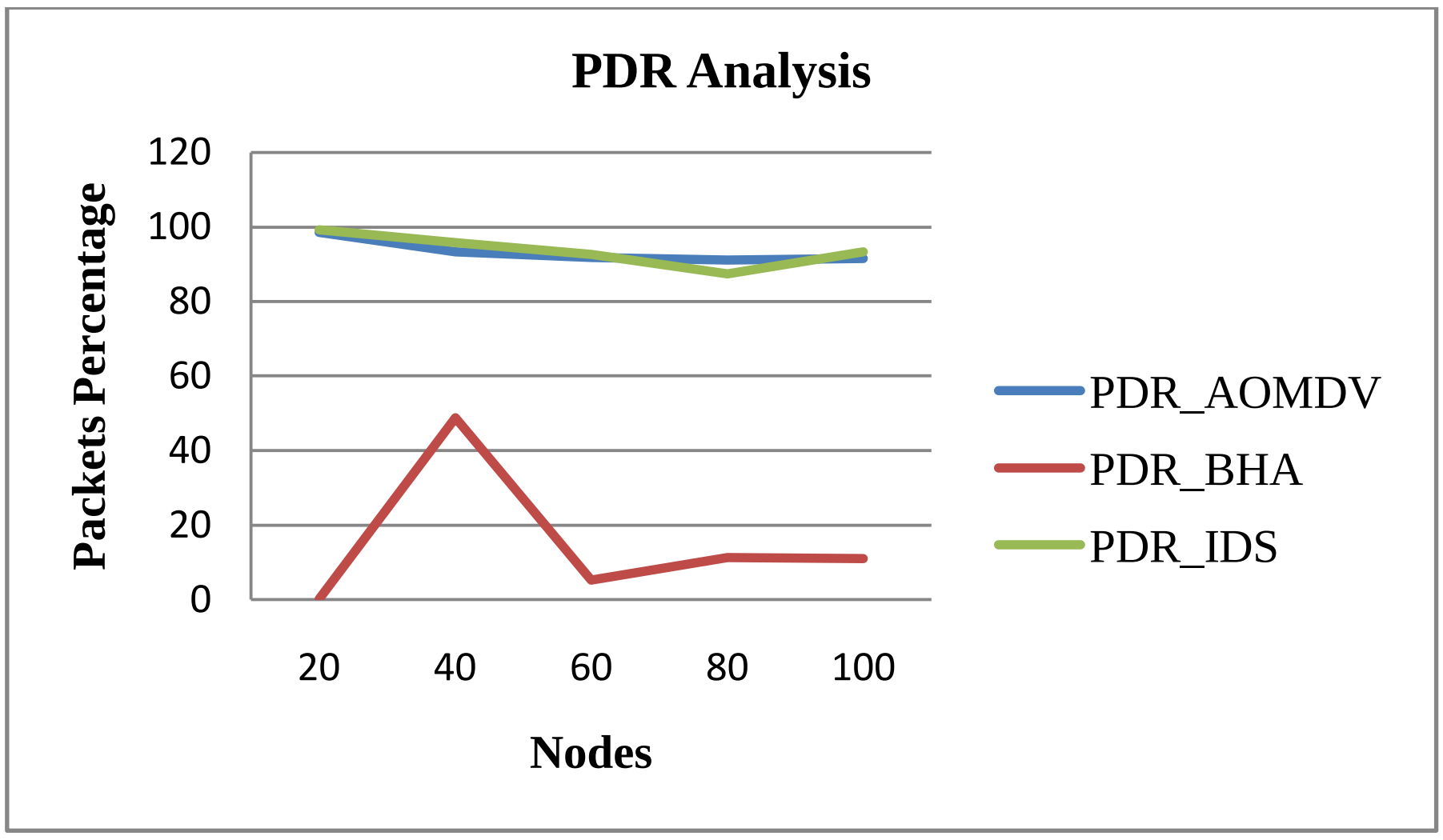

**Fig. 2 PDR Analysis**

## c). Routing Overhead Analysis in case of AOMDV, Attack and IDS

The routing overhead is counted through the number of routing packets are deliver in network. The routing packets are flooding in network to establishment connection in between sender and receiver through intermediate nodes. The nodes are forming dynamic topology by that the link establishment is the challenging issue in MANET. This graph illustrated the routing overhead in case of AOMDV, Attack and IDS scheme and observes that the performance of IDS scheme is recovers the performance in presence of attacker in 20, 40, 60, 80 and 100 nodes scenario. The routing overhead of 20 nodes is about 315 and rest of them is more than 20 except 40 nodes. In case of IDS the packet receiving is more as respect to routing packets are deliver in network but in case of attacker it is negligible compare to that. The routing overhead of normal and IDS are overlapped due to that not visible clearly and it is too much less then attacker performance.

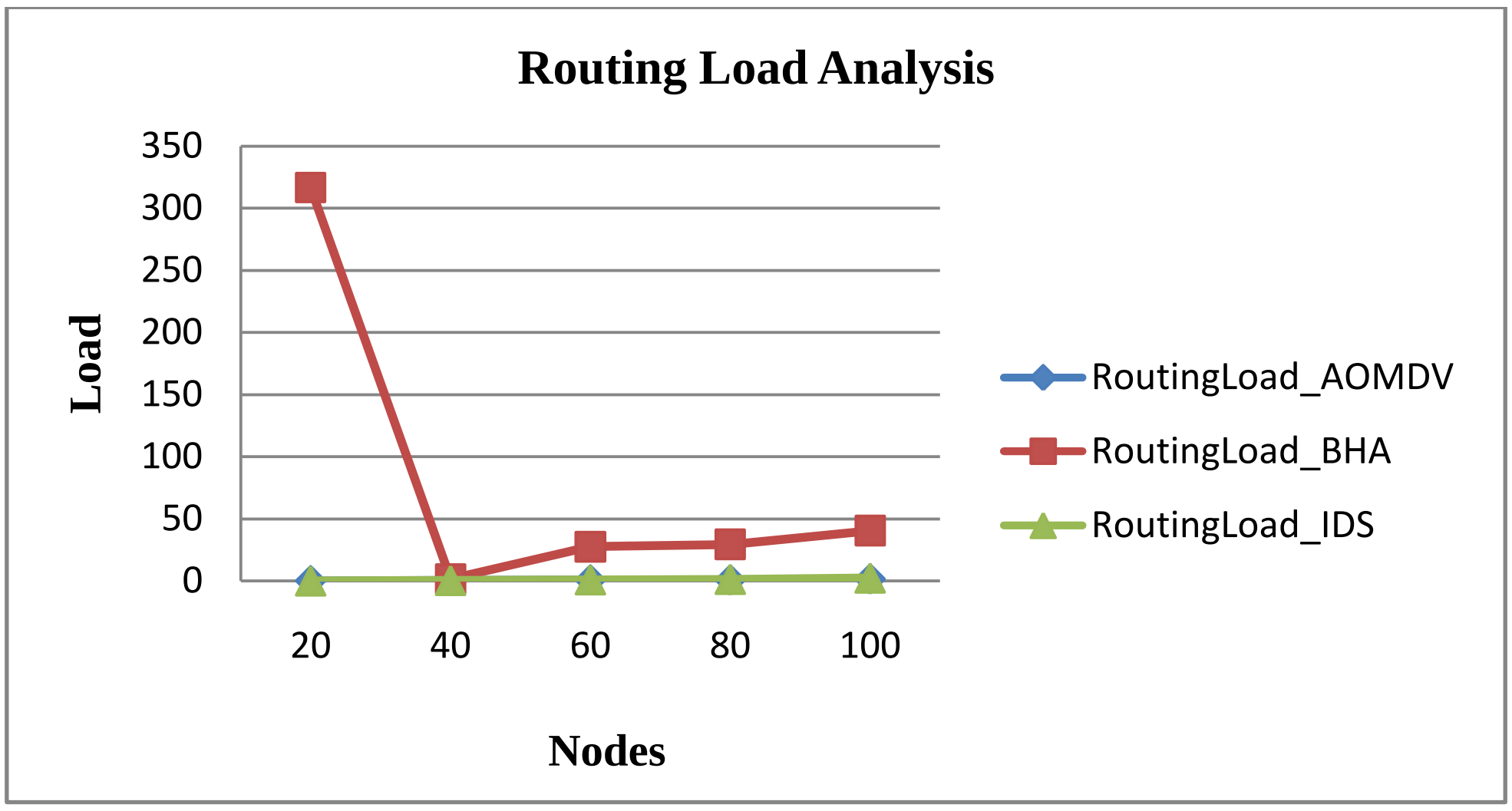

**Fig. 3 Routing Load Analysis**

## d). Throughput Analysis in case of AOMDV, Attack and IDS

The packets receiving in MANET is not being on any supervision or administrator. The data delivery in that kind of network is not safe. In this graph we illustrated the throughput analysis in case of normal AOMDV, Attack and proposed IDS scheme. The packet per unit of time in case of attack is almost negligible in network but in case of proposed IDS scheme the throughput is much better as compare to attacker in 20, 40, 60, 80 and 100 nodes scenario. The throughput in case of normal AOMDV routing is about more than 3500 packets/seconds and not less than 600packets/seconds. It implies that the throughput in case of IDS is more as compare to normal AOMDV routing. The reason behind is that if the attacker is existing in established path then in that case that path is not select for data delivery to maintaining the reliability and the next alternative path is chosen more reliable and strong that minimizes packet dropping and improves data delivery in presence of attacker.

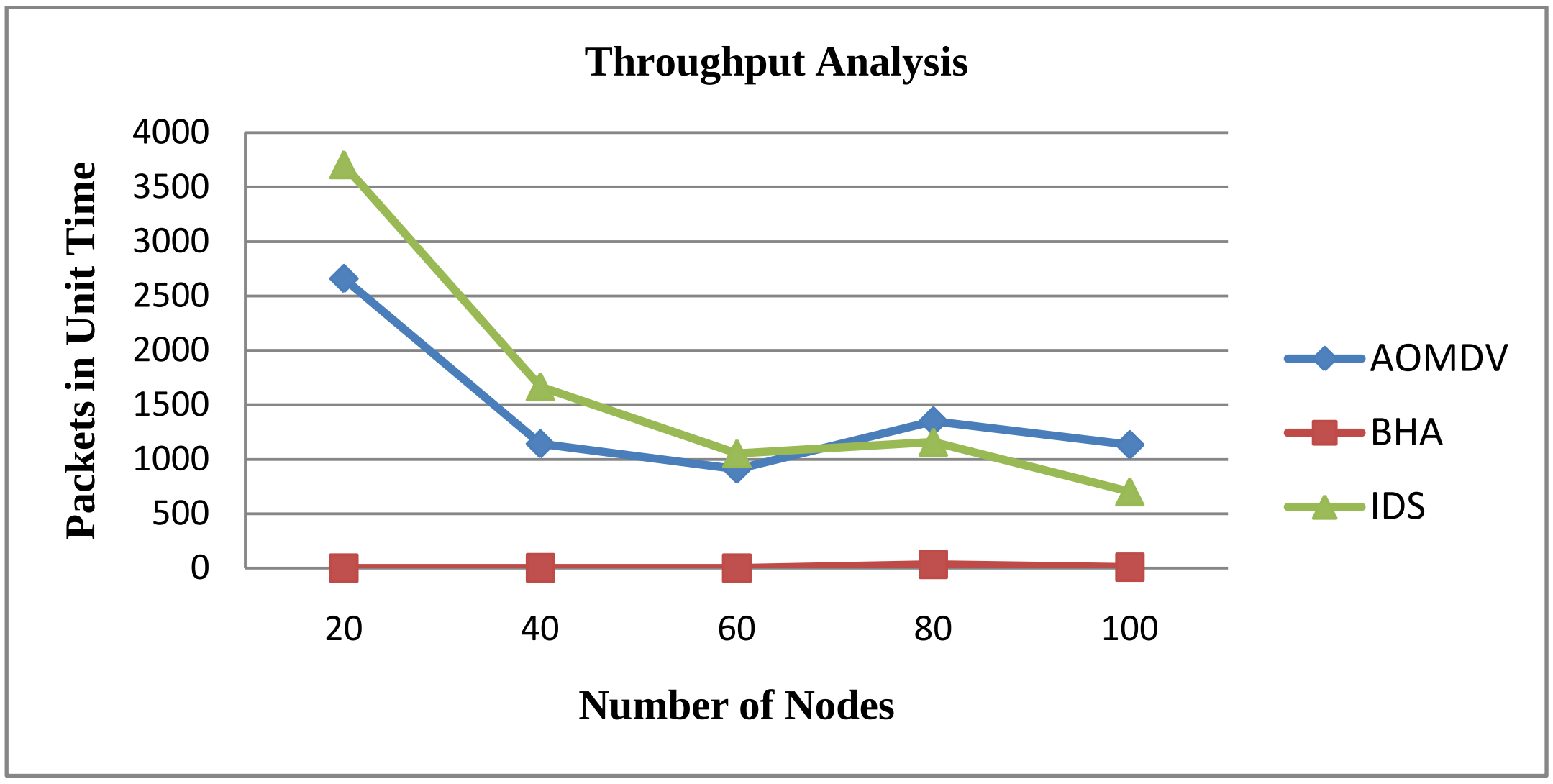

**Fig. 4 Throughput Analysis**

### e). Packets Receive Analysis in case of AOMDV, Attack and IDS

Multipath routing is enhanced the possibility of successful receiving by proving the alternative path in network if first one is fail. The UDP protocol is the transport layer protocol provides end to end delivery in network. In this graph the performance of UDP packet received is examine in case of normal AOMDV, Attack in AOMDV and proposed IDS scheme. The packet receiving in case attacker is negligible due to routing misbehavior of AOMDV protocol. The packet receiving in case of proposed IDS scheme is much more in 20, 40, 60 80 and 100 nodes densities also AOMDV provides the almost same performance. That provides the better receiving.

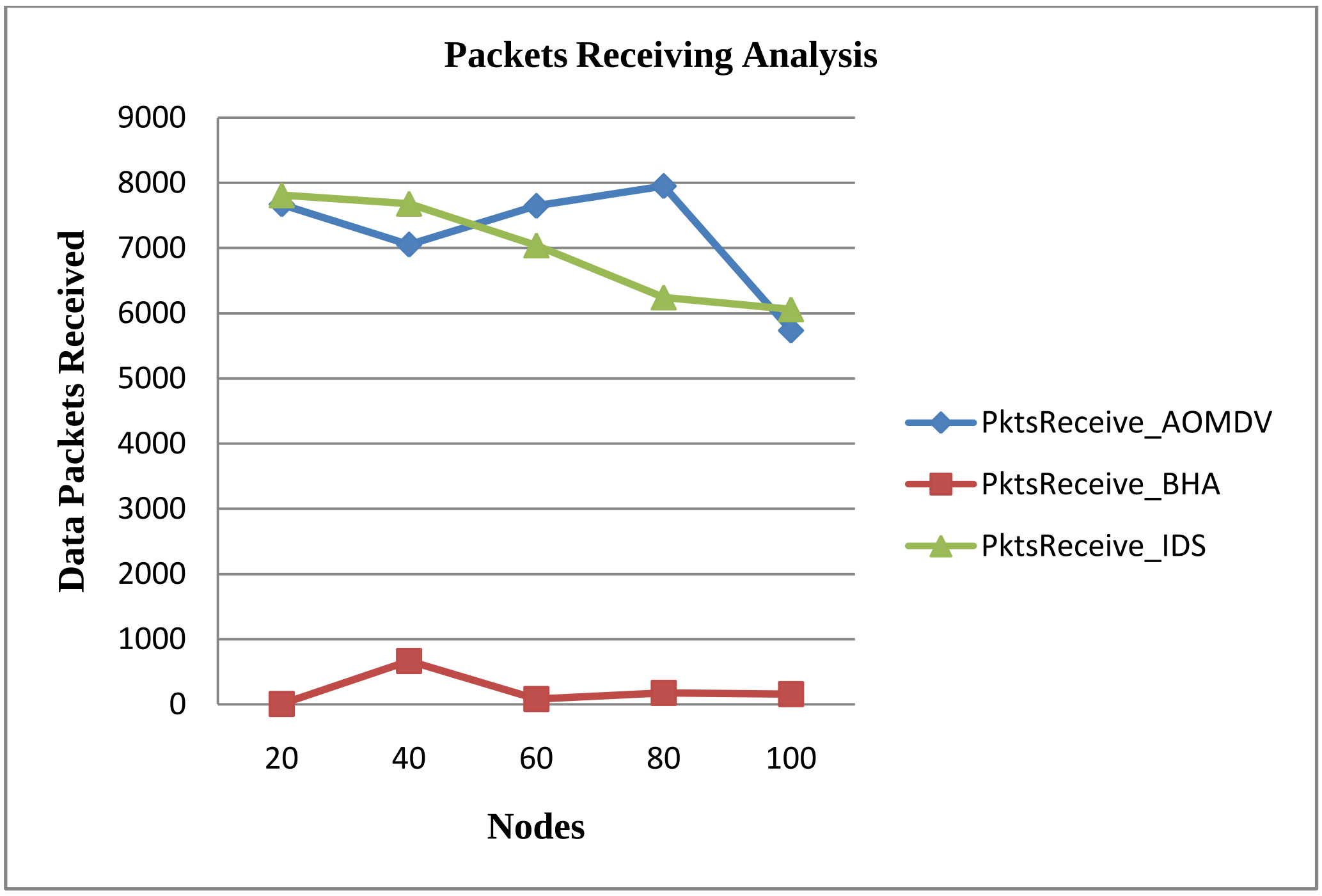

**Fig. 5 Packets Received analysis**

## VI. Conclusion & Future Work

The central coordination system absence, security is the major issue in MANET. The data packets in network are delivering in between sender and receiver through routing mechanism of connection establishment. The performance is illustrated in 20, 40, 60, 80 and 100 nodes scenario. The attackers are dropping the all data packets that are the reason of routing misbehavior in MANET. The malicious attacker action is wedged by proposed IDS security scheme and provides the attacker free network. The AOMDV protocol provides the alternative if the problem in accessible path is occurred. The routing performance is measured by performance metrics in case of normal AOMDV routing, Malicious Attack and proposed IDS scheme. The proposed IDS scheme identified the attacker through next hop information of data delivery and also forward the Identification of node ID of attacker in network. If that ID is exist in routing establishment then the alternative route is select for data delivery.

The routing performance of AOMDV protocol and IDS scheme on AOMDV is almost equal that means nearly the network is provides equivalent performance. In attacker module degrades the whole performance of network but in presence of attacker their activities are completely blocked by IDS scheme after identifying them in network. Moreover after dump the performance of network by attacker proposed IDS scheme recovers 95 % of data loss as compare to normal AOMDV.

In future we also apply this IDS scheme on other routing attacks like wormhole attack and Grey-hole attack. Also analyze the effect of attack on energy consumption of mobile nodes i.e. the major or only source of communication. Without energy existence nodes in MANET are not survives for a long time.

## Acknowledgement

I would like to thank Prof. Anurag Jain, Assistant Prof. Chetan Agrawal, for accepting me to work under his valuable guidance. He closely supervises the work over the past few months and advised many innovative ideas, helpful suggestion, valuable advice and support.